# Resilience in PON-based data centre architectures with two-tier cascaded AWGRs

Mohammed Alharthi[1], Sanaa H. Mohamed[2], Taisir E. H. El-Gorashi[2], Jaafar M. H. Elmirghani[2]
[1]*School of Electronic and Electrical Engineering, University of Leeds, LS2 9JT, United Kingdom*
[2]*Department of Engineering, King's College London, WC2R 2LS, United Kingdom*
elmaalh@leeds.ac.uk, sanaa.mohamed@ kcl.ac.uk, taisir.elgorashi@kcl.ac.uk, jaafar.elmirghani@kcl.ac.uk

**ABSTRACT**
This paper investigates the performance of a two-tier AWGR-based Passive Optical Network (PON) data centre architecture against an AWGR-based PON data centre architecture by considering various scenarios involving link failures to evaluate the resilience of both designs. To optimize traffic routing under different failure scenarios, a Mixed Integer Linear Programming (MILP) model is developed and the power consumption and delay performance is assessed. The results demonstrate that the two-tier AWGR architecture reduced the power consumption and the delay compared to the AWGR-based architecture by up to 10% and 61%, respectively.

## 1. INTRODUCTION

In recent years, there has been a significant increase in the volume of traffic requiring processing and transportation, largely attributed to the rapid adoption of Internet-based applications [1]. Researchers have been dedicating efforts towards optimizing access and core communication network designs [2]-[8], as well as developing energy-efficient data centres [9]-[15] to effectively handle the increasing demands of Internet traffic while ensuring energy efficiency. Current data centre architectures face significant limitations and challenges including high costs, high latency, low throughput, complex management requirements, and limited scalability [16]-[18]. One of the research directions for addressing these challenges focused on integrating Passive Optical Network (PON) technology into Data Centre Networks (DCNs) [19]-[22]. Studies have shown that PON technology can provide high capacity, cost-effectiveness, elasticity, scalability, and energy efficiency in future data centre designs [9]. For inter-rack connectivity, different techniques including Wavelength Division Multiplexing (WDM) PON, Orthogonal Frequency Division Multiplexing (OFDM) PON, and Arrayed Waveguide Grating Routers (AWGRs), have been employed in PON-based DCNs [20]-[23]. To facilitate communication between servers within the same rack, passive devices such as Passive Polymer Backplanes, Fibre Bragg Gratings (FBGs), and passive star reflectors have been proposed in [9]. Novel AWGR-based PON data centre architecture, that provide WDM connectivity by using AWGRs, have been introduced and studied in [15], [24]. This design, referred to as PON 3, contains four racks where each rack is supported by a single splitter and a single coupler, two $M \times M$ AWGRs, and a single OLT. In [25]-[27], we proposed a two-tier cascaded-AWGRs data centre architecture that comprises four cells where each cell consists of four racks interconnected through two couplers and two splitters to the rest of the data centre. The two tiers of AWGRs connect the cells to four OLT switches, which facilitate multipath routing, reduce oversubscription, and enhance load balancing.

Data centres must possess resilience as any period of inactivity can disrupt the flow of information, leading to substantial financial repercussions. Network resilience pertains to the network's capacity to maintain stable operation even in the event of failure [28]. Failures within DCNs can occur due to equipment malfunctions, severed connections, natural calamities, and cyber-attacks [29], [30]. Enhancing the resilience of DCNs can be accomplished through several strategies, such as network redundancy, diversity, traffic restoration, and prevention [29], [31], [38]. Prevention involves situating network components in secure locations and furnishing them with backup power supplies. Traffic restoration entails redirecting demands to mitigate the impact of failures. Redundancy and diversity are established in networks by replicating network components to serve as backups in case of failure. These redundancies and diversities can be implemented at different levels, including equipment and network pathways. For instance, the addition of multiple network interfaces to network nodes constitutes redundancy at the equipment level. This approach is employed when the likelihood of multiple simultaneous failures of links and nodes is low. In the network pathways redundancy scheme, paths are designated to be both link and node disjoint [29].

This paper compares the performance of the proposed two-tier cascaded-AWGRs data centre architecture to PON 3 data centre architecture under different links failure scenarios to investigate their resilience and to evaluate the impact of the failures on the power consumption and the delay. This study aims to determine which AWGR-based PON data centre architectures exhibit greater resilience. In order to compare the performance, a Mixed Integer Linear Programming (MILP) model is developed to optimize traffic routing under various failure scenarios. The reminder of this paper is organized as follows: Section 2 describes the system model for the proposed design and the system model for PON 3. Also, this section discusses the types of considered failures.

Section 3 describes the developed MILP optimization model. Section 4 discusses the results, and finally, Section 5 provides the conclusions of this paper.

## 2. SYSTEM MODEL AND CONSIDERED TYPES OF FAILURES

In PON3, as depicted in Figure 1, the servers within each rack are linked to a coupler, which in turn is connected to one of the input ports of the AWGRs for the uplink. Two 4 × 4 AWGRs are employed to facilitate full interconnection between racks, as well as between racks and a single OLT switch. For the downlink, one of the AWGR output ports is linked to a splitter, which then connects to servers within a rack. In contrast, the proposed architecture, as depicted in Figure 2, introduces a two-tier cascaded level of AWGRs, enabling complete interconnection between the cells and between cells and different OLT switches. Each server in the proposed architecture is equipped with two uplinks, where a coupler per uplink is linked to an input port of a different AWGR in the first level of the 2-tier cascaded AWGRs fabric. For the downlink, two different output ports of two different AWGRs at the first level are used where each is connected to a splitter, which in turn connects to servers within a cell for one of the downlinks. The connectivity approach of PON 3, as depicted in Figure 1, relies on single-path interconnection between communicating pairs, raising concerns regarding connectivity in the event of a link failure. Such failures could potentially render all servers in a rack inaccessible due to the single-path connectivity between this rack to other racks. In contrast, the proposed two-tier AWGR-based architecture, illustrated in Figure 2, employs duplicated links, facilitating multi-path routing and enhancing connectivity compared to traditional PON-based architectures. Moreover, it addresses limitations such as resilience and load balancing capabilities.

We consider five types of link failures in this work which are the failure of a link between: (i) a server and a coupler (F1), (ii) a coupler and an AWGR (F2), (iii) two AWGRs (F3), (iv) an AWGR and a splitter (F4), and (v) a splitter and a server (F5). In PON 3, all the servers in a rack could become inaccessible because of a single link failure in the single path that connects this rack to the rest of the architecture. Therefore, link failures F1, F2, F4 and F5 in PON 3 will result in totally disconnecting some servers. PON 3 architecture can survive multiple F3 failures (failure of a link between two AWGRs) at the cost of increased power consumption due to utilising the OLT. On the other hand, the multipath routing of the proposed two-tier AWGRs architecture makes the architecture resilient against all the considered failures while avoiding the use of the OLT.

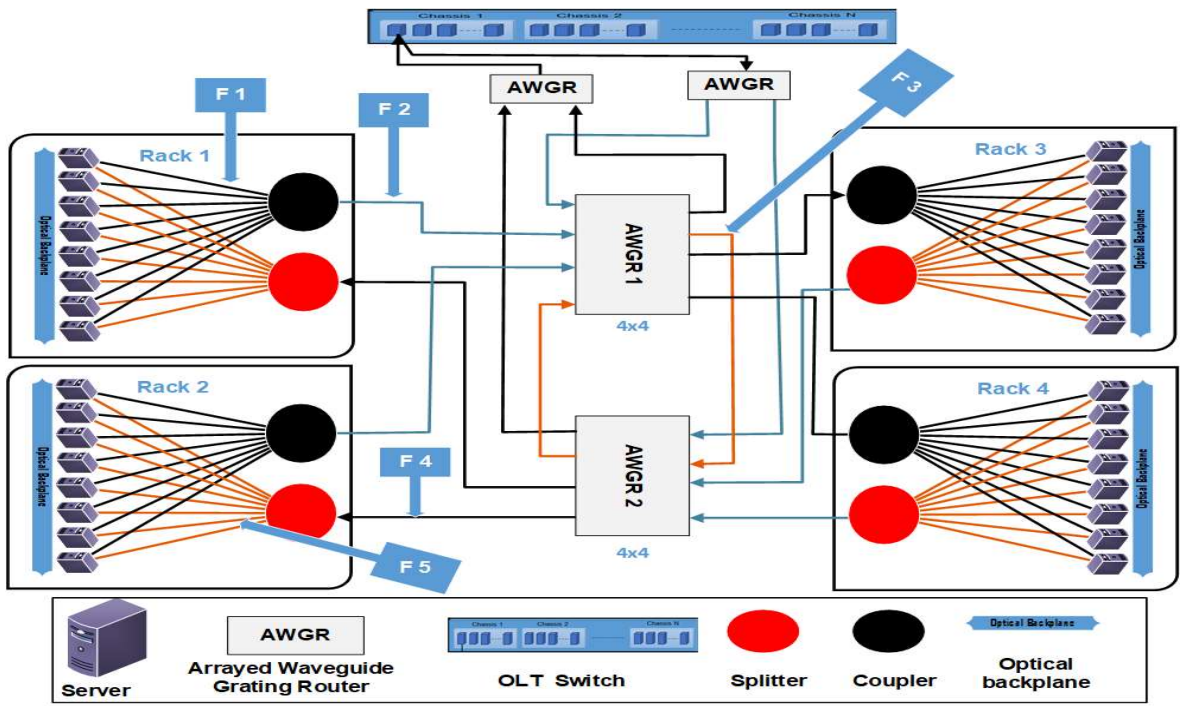

*Figure 1. The AWGR-based data centre architecture (PON 3) (Each of the labels F1, F2, F3, F4, and F5 provides an example of a failure type).*

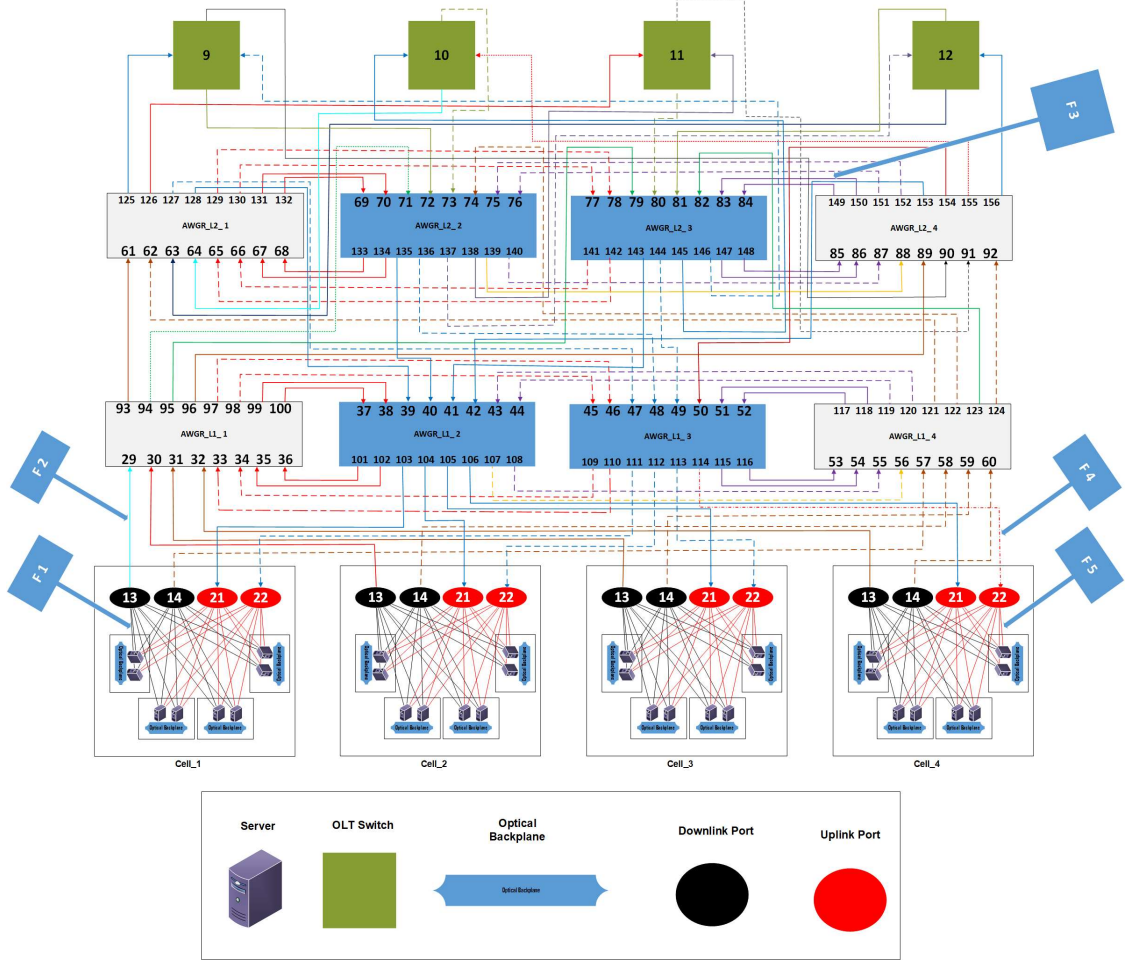

*Figure 2. The-two tier cascaded-AWGRs architecture with for cells (Each of the labels F1, F2, F3, F4, and F5 provides an example of a failure type).*

## 3. MILP Model for Energy-Aware Routing Under Link Failures in the Proposed and PON 3 Architectures

This section describes the MILP optimization model we developed to achieve energy-aware routing over the proposed design and the PON 3 design, when considering different failure scenarios. The objective of the MILP model is to minimize the total power consumption for both designs. The model adheres to all routing restrictions, capacities and utilization constraints, and queuing delay evaluation constraints in order to meet the objective.

$$\textit{Minimize:} \quad P = \sum_{s \in S} \vartheta_s + \sum_{t \in T} \mu_t \,, \quad (1)$$

where $P$ refers to the total networking power consumption in the architecture, $\vartheta_s$ indicates the total power consumption of physical servers which is composed of the idle power consumption of servers, the proportional power that is a function of the CPU usage as a result of processing data and receiving and transmitting traffic, and the ONUs power consumption that is used for communication. Finally, $\mu_t$ refers to the total power consumption of OLT switch which is composed of the idle power of the OLT switch and the proportional power that represents the power used to forward traffic.

## 4. RESULTS AND DISCUSSION

The parameters used to evaluate the power consumption and delay are shown in Table 1. Figure 3 shows the comparison between the proposed architecture and PON 3 architecture in terms of the power consumption under different types of link failure. Under the normal mode (NF), the total power consumption of both architectures is equal under the same traffic load as the same number of servers and OLT ports are activated.

*Table 1. Parameters for the MILP model.*

| Parameter | Value |
|---|---|
| Server's maximum power consumption | 301 W [33] |
| Server's idle power consumption | 201 W [33] |
| Data rate of servers | 1 Gbps |
| Power consumption of ONU | 2.5 W [34] |
| Data rate of ONU | 10 Gbps |
| OLT switch maximum power consumption | 1940 W [35] |
| OLT switch idle power consumption | 60 W [35] |
| Data rate of OLT Switch | 8600 Gbps [35] |
| Traffic demand between nodes | Random and uniformly distributed, 0.2 – 0.8 Gbps |
| Physical link capacity | 10 Gbps |
| Typical examples of server traffic load and resulting delay using an M/M/1 queuing | [200, 400, 600, 800] (Mb/s) = [15, 20, 30.1, 60.2] (µ sec/packet), respectively |
| Typical examples of OLT traffic load and resulting delay using an M/M/1 queuing | [200, 400, 600, 800] (Mb/s) = [0.00139538, 0.00139544, 0.00139545, 0.00139547, 0.00139551] (µ sec/packet), respectively. |

Regarding the link failure types F1 (link failure between a server and a coupler), F2 (link failure between a coupler and an AWGR), F4 (link failure between an AWGR and a splitter) and F5 (link failure between a splitter and a server), some servers will be disconnected as explained in Section 2 and therefore we do not show the power consumption values. However, the proposed architecture can handle these types of failures due to its multipath routing capabilities while maintaining the same level of power consumption as normal mode operation.

Under link failure F3, the total power consumption of the proposed architecture is less than the total power consumption of PON 3 architecture by up to 10%. This is because PON 3 design must relay traffic through an OLT switch which adds to the power consumption. For the proposed architecture, the availability of more links between the AWGR's in the two tiers leads to using the alternative link instead of relying on an OLT switch.

Figure 4 shows the queuing delay under the various types of link failures in the proposed architecture and in PON 3 architecture. For both designs under the normal mode (NF), the queuing delay is equal as the same number of servers and OLT ports are activated. Note that the results are generated when considering the same traffic load for both architectures. The proposed architecture, which can handle F1, F2, F4, and F5, maintains the same value of queueing delay compared to the value of queueing delay in normal mode (NF). Note that this is the delay experienced at the source and destination servers. Under link failures F3, the total queuing delay in PON 3 architecture increases by up to 61% as a result of relaying the traffic through the OLT. In contrast, the queuing delay remains the same as it was in the normal mode in the proposed architecture because of the use of alternative routes through the AWGRs without the need to go through the OLT.

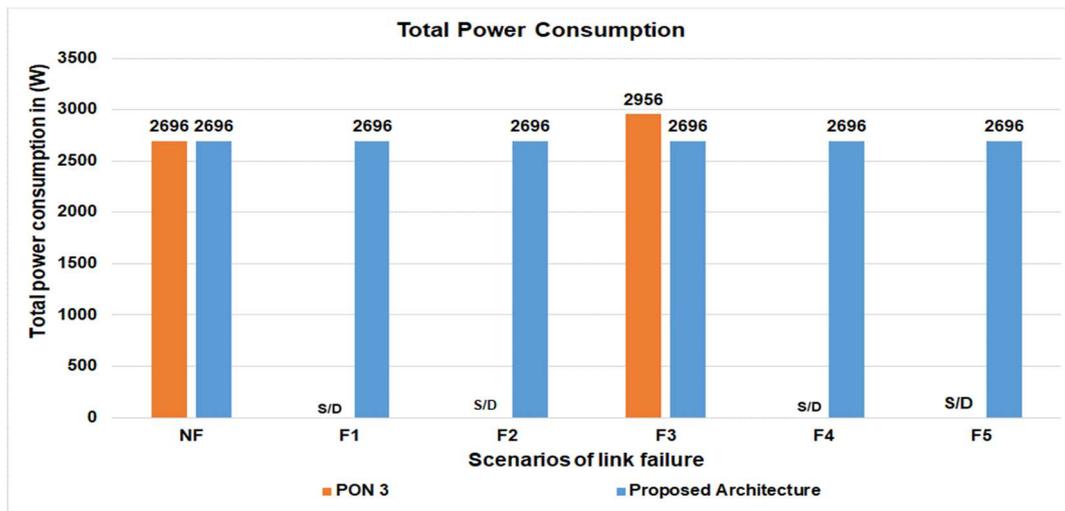

*Figure 3. Comparison of the total power consumption between PON 3 and the proposed architecture under various failure scenarios (S/D stands for a system is down).*

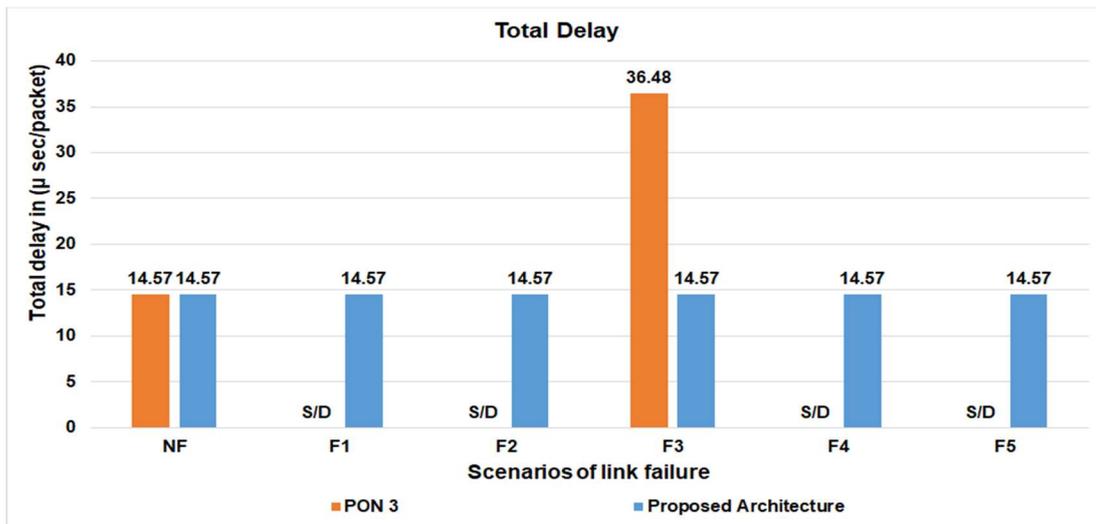

*Figure 4. Comparison of the total delay between PON 3 and the proposed architecture under various failure scenarios (S/D stands for a system is down).*

## 5. CONCLUSIONS

This paper presented an investigation into the performance of the proposed two-tier cascaded-AWGRs data centre architecture in comparison to the PON 3 data centre architecture under link failures. Various scenarios involving link failures were examined to assess resilience in both designs. Performance evaluation was conducted employing two metrics: power consumption and delay. A MILP optimization model was developed to assess the performance of the proposed two-tier cascaded-AWGRs data centre architecture against the PON 3 architecture under different failure scenarios. The findings indicated that the proposed data centre architecture reduced the power consumption and the delay compared to the PON 3 architecture and demonstrated greater resilience across various link failure scenarios.

## ACKNOWLEDGEMENTS

The authors would like to acknowledge funding from the Engineering and Physical Sciences Research Council (EPSRC), INTERNET (EP/H040536/1), STAR (EP/K016873/1) and TOWS (EP/S016570/1) projects. All data are provided in full in the results section of this paper. The first author would like to thank the Ministry of Interior (MOI), Saudi Arabia for funding his PhD scholarship.